# Alternative sum rules and waterbed effects of Lorentz resonator system for sound absorption and transmission in a unidimensional waveguide

Di Mo[a,b], Yumin Zhang[a,b*], Tianquan Tang[a,b], Xiaochao Ji[a,b*], Xiang Liu[a,b], Keming Wu[c]

[a] *School of Mechatronic Engineering and Automation, Foshan University, Foshan 528000, China*

[b] *Guangdong Provincial Key Laboratory of Industrial Intelligent Inspection Technology, Foshan University, Foshan 528000, China*

[c] *College of Underwater Acoustic Engineering, Harbin Engineering University, Harbin, China*

Corresponding to Yumin Zhang yuminzhang@fosu.edu.cn

## Abstract

We investigate fundamental constraints on passive linear time-invariant acoustic systems through the developing alternative linear sum rules for sound absorption and transmission. Our approach, based on the Herglotz function method, yields integral identities without non-linear logarithmic terms or frequency weightings, providing clearer physical insights into system performance limits. The study focuses on unidimensional waveguides with Lorentz resonators, encompassing various practical acoustic structures. The developed sum rules are found to be particularly effective in predicting constraints on the average sound absorption coefficient for broadband absorbers operating in deep-subwavelength structures. Based on these rules, we demonstrate the waterbed effect in such systems, highlighting the inherent compromises between absorption efficiency, bandwidth, and device thickness. Through case studies of resonator arrays and membranes, we illustrate the practical implications of these new sum rules for designing optimal sound absorbers and isolators. The work concludes with a discussion on the challenges and future prospects in passive noise control, suggesting potential pathways to surpass current performance boundaries.

# 1. Introduction

Sum rules impose fundamental constrains for casual, passive, and linear time-invariant acoustics systems [1,2]. A sum rule often takes the form of $\int_0^\infty \omega^{-n} \ln T(\omega)\, d\omega = C$, where $C$ is a constant. To The best of the authors' knowledge, the sum rule can be traced back to Bode's sensitivity integral [3,4] in control theory, which is $\int_0^\infty \ln|S(\omega)|\, d\omega = 0$ for feedback control system, where $S(\omega)$ is normal called the sensitivity function. It's derived from Cauchy's contour integral theorem, which requires $S(\omega)$ a rational function with two more poles than zeros. The Bode's sensitivity integral was later interpreted as well-known waterbed effect [5] and conservation of dirt [6]. Fano [7] further developed an integral relation showing the compromise between the bandwidth and the logarithm sensitivity of reflection for impedance matching of circuit networks, which now is termed as Bode-Fano bound. Cauchy's integral theorem has also been extensively used in electromagnetism to identify physical limitations in antenna design [8,9], wave scattering [10], radar wave absorption [11], and other applications. Bode's gain-phase integral, which is essentially identical to the Hilbert-Huang transform and the Kramers-Kronig relations, is also widely employed to derive fundamental limitations across various fields of physics [12], including ultrasound absorption and acoustic scattering [13,14], optics [15], and quantum chromodynamics [16]. These theories can be directly applied to one-dimensional acoustic network analysis. Yang et al. introduced an integral identity for reflection sensitivity [2], which provides the ultimate thickness of a sound absorber required to achieve an arbitrary absorption band in a one-dimensional waveguide, assuming the absorption function is rational. Bravo and Maury further established absorption bounds for micro-perforated panels [17,18].

For non-rational transfer functions, such as confluent hypergeometric and Bessel functions frequently used in acoustic cavity and membrane analysis, it is more appropriate to adopt the Herglotz function method [1,19] (also known as the Nevanlinna function in many contexts) to construct integral identities and derive the sum rules.

The Herglotz function method operates in Stolz domain. In complex analysis, for a complex frequency $\omega$, the domain is $\varphi \leq \arg \omega \leq \pi - \varphi$ for $\varphi \in (0, \pi/2]$ [19]. It means that all operations take place in complex half plane excluding real axis. With Herglotz function method, we only need to know the static ($|\omega| \to 0$) and dynamic properties ($|\omega| \to \infty$), without considering the specific designs or intricate band structures, to derive the sum rules of a passive LTI system. S These unique properties make the method a versatile mathematical tool for establishing fundamental limitations in many physical systems, such as cloaking [20], optical scattering of periodical structure [21,22]. A recent review [12] thoroughly summarized the application of sum rules across various methods in multiple areas of physics. In acoustics, Meng et al. applied the mathematical framework of the Herglotz function developed in [19] to derive sum rules for sound scattering in 1D waveguides, both with and without flow [1,23]. The significant results include integrals for weighted logarithm reflection sensitivity (eg., $\omega^{-n} \ln R$) and transmission loss (eg., $\omega^{-n} TL$) o porous mediums and side-branch liners, all of which are mainly determined by the depth of the effective medium. With considering viscothermal losses, it's found that the adiabatic index $\gamma$ of air must be taken into account for lossy medium such as porous materials.

The sum rules provided in [1,2,14,17,18,23] impose fundamental constrains on noise treatments and the manipulating sound waves through passive measures, explicitly demonstrating what can and cannot be achieved with passive LTI systems. However, existing sum rules for noise control by passive LTI system are in the form of $-\int_0^\infty \omega^{-2} \ln|T(\omega)| = C$, $T(\omega)$ is reflection and transmission coefficients. The integrand contains two weighting functions, the logarithm operator and $\omega^{-2}$ leading that its physical interpretation from an energy perspective is not always immediately clear. In this work, we revisit the theories in [1,19], and using Herglotz function method to give alternative sum rules for passive LTI system consisting of Lorentz resonators. The sum rules, without logarithm operators, are obtained through integrating sound absorption coefficient and sound transmission coefficients directly. Additionally, sum rules without weighting functions are presented. Based on these, we elaborate on the waterbed effect for sound absorption and transmission, following the intuitive expressions from control theory.

In the following sections, Section 2 describes the problems and the method used to address which, Sections 3 presents the sum rules for sound absorption with typical cases, Section 4 gives sum rules for sound transmission problem with cases studies, and the final section provides discussions and conclusions.

## 2. Problems and methods

### 2.1 Sound absorption and transmission problems

In this work, we examine sound absorption and transmission issues in a one-dimensional (1D) waveguide, as depicted in Fig. 1. The structures under study consist of Lorentz resonators. The surface of the structure is indicated by the solid line in the 1D waveguide shown in Fig. 1. The real structures could include Helmholtz resonators, membranes, suspended diaphragms, and their composites arranged in series, parallel, or combined configurations. Note that we only consider the case of plane wave incidence, allowing us to neglect high-order modes and treat subwavelength structures with lumped parameters.

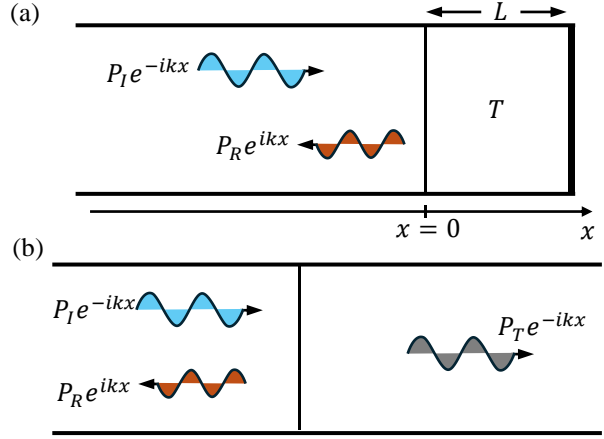

Figure 1: 1D model for sound absorption and transmission problems of passive Lorentz resonators. (a) 1D model for sound absorption with rigid backing. (b) 1D model for sound transmission problem.

The pressure $P = p/\rho_0 c_0^2$ and particle velocity $U = u/c_0$, as labeled in Fig. 1, are normalized values, where $c_0$ represents the speed of sound and $\rho_0$ denotes the density of air. The normalized acoustic impedance, denoted by $Z$, is the specific impedance normalized by $Z_0 = \rho_0 c_0$. For a fundamental unit of a Lorentz resonator, the acoustic impedance is given by $Z = Mi\omega + D + K i\omega - i\cot kL$, where $M$, $D$, and $K$ are the dynamic mass, resistance, and stiffness per unit area normalized by $Z_0$, respectively. Here, $\omega$ is the complex frequency and $k = \omega/c_0$ is the complex wavenumber.

The definition of passivity follows [1,19], which would not be repeated here. Fundamental units forming the system could be micro-perforated panels [24], Helmholtz resonators [25], stretched membranes [26], passive shunted loudspeakers [27,28] and passive shunted PZT plates [29] and their hybrids. In addressing sum rules for sound absorption, we examine the entire frequency and wavelength spectrum to determine the portion of incident energy that can be absorbed by the terminating structure. For sound transmission, the objective is to quantify the fraction of energy blocked by the partition. This analysis involves integrating sound absorption and transmission coefficients over the full range of frequencies (or wavelengths) from zero to infinity. Previous studies [1,2] have approached these integrals using integrands in the form of logarithmic sensitivity, weighted by the negative square of frequency. In contrast, our work presents integral identities and inequalities that exclude the logarithmic operator and incorporate both weighted and unweighted forms, offering an energy-centered perspective. We use the Herglotz function method to achieve our objectives. For ease of understanding, this method is summarized in the following subsection.

**2.2 Herglotz functions method**

A function that maps the complex upper half-plane onto itself or onto a real constant is often referred to as a Herglotz function. This mapping serves as a powerful tool for analyzing passive systems. From an engineering perspective, it can be viewed as an integral representation that converts an integration problem to one of finding static and dynamic limits. This mapping method allows us to derive identities that relate weighted integrals of a Herglotz function to its asymptotic expansion, that's

$$\lim_{\varepsilon \to 0^+} \lim_{\delta \to 0^+} \frac{2}{\pi} \int_{\varepsilon}^{\varepsilon^{-1}} \frac{Im[H(\omega + i\delta)]}{\omega^{2q}} d\omega = a_{2q-1} - b_{2q-1}, q = -(M-1), \dots N \quad (1)$$

where $a_{2q-1}$ and $b_{2q-1}$ are specific coefficients in the asymptotic expansion of $H(\omega)$ in its static and dynamic limits:

$$H(\omega)|_{\omega \hat{\to} 0} = a_{-1}\omega^{-1} + a_1\omega + \cdots + a_n\omega^n + \cdots a_{2N-1}\omega^{2N-1} + o(\omega^{2N-1}), \quad (2a)$$

and

$$H(\omega)|_{\omega \hat{\to} \infty} = b_1\omega + b_{-1}\omega^{-1} + \cdots + a_m\omega^m + \cdots + b_{-(2M-1)}\omega^{-(2M-1)} + o(\omega^{-(2M-1)}). \quad (2b)$$

$a_n$, $b_m$ are real coefficients, and $M$ and $N$ are non-negative integers. The notation $\hat{\to}$ denotes $|\omega| \to \infty$ in the Stoltz domain, which is $\theta \leq \arg \omega \leq \pi - \Theta$, for any $0 < \Theta < \pi/2$ as shown in Fig. 2.

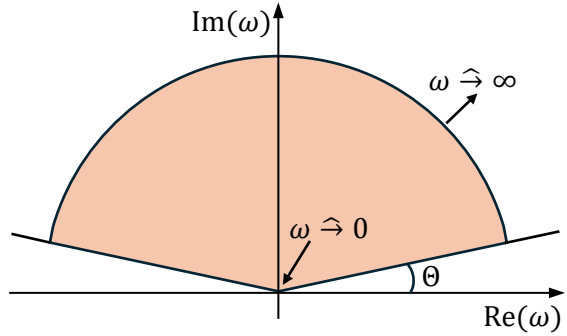

Figure 2: Stolz domain.

In the Stolz domain, all mathematical operations are carried out in the complex $\omega$ plane rather than on the real axis. It is important to note that the functions used to construct Herglotz functions $H(\omega)$ must be holomorphic. Specifically, in this work, the constructed Herglotz functions are symmetrical, allowing the left-upper half of the complex plane to map onto the right-upper half [19].

With different $q$ values in Eq. (1), sum rules with a different weighting term $\omega^{2p}$ in Eq. (1) can be constructed. Note that not all $p$ values give the results of above integral. For example, the $a_{2q-1}$ or $b_{2q-1}$ don't exist in the asymptotic expansions. In present work, we take $q = 0$ and $q = 1$ to obtain the sum rules for sound absorption and transmission problems, and with which we get two integral identities:

$$\lim_{\varepsilon \to 0^+} \lim_{\delta \to 0^+} \frac{2}{\pi} \int_{\varepsilon}^{\varepsilon^{-1}} \frac{Im[H(\omega + i\delta)]}{\omega^0} d\omega = a_{-1} - b_{-1}, \quad (3a)$$

and

$$\lim_{\varepsilon \to 0^+} \lim_{\delta \to 0^+} \frac{2}{\pi} \int_{\varepsilon}^{\varepsilon^{-1}} \frac{Im[H(\omega + i\delta)]}{\omega^2} d\omega = a_1 - b_1, \quad (3b)$$

where $a_{-1}$ and $a_1$ are the static limits for $q = 0$ and $q = 1$, respectively, and $b_{-1}$ and $b_1$ is the dynamic limits. Using the integral representations above, we can construct Herglotz functions related to sound absorption and transmission functions to derive corresponding sum rules.

## 3. Linear sum rules for sound absorption

### 3.1 Derivation of sum rules

In this section, we develop alternative sum rules using the Herglotz function method, specifically focusing on the integral of $\alpha(\omega)$ without using logarithmic operators or frequency weighting functions. The challenge lies in constructing proper Herglotz functions; while the solution may seem straightforward once found, the search itself often appears daunting. The absorption coefficient $\alpha(\omega)$ for normal incidence in a 1D waveguide is given by

$$a(\omega) = 1 - |R(\omega)|^2 = 1 - \left|\frac{Z(\omega) - 1}{Z(\omega) + 1}\right|^2 = \frac{4\theta(\omega)}{[1 + \theta(\omega)]^2 + \chi^2}, \qquad (4)$$

where $\theta = \text{Re}(Z)$ and $\chi = \text{Im}(Z)$ are real and imaginary parts of the normalized acoustic impedance $Z(\omega)$, and $R$ is the complex pressure reflection coefficient. Eq. (1) contains an absolute value operator, which is not a holomorphic function and thus cannot be used to construct a Herglotz function for deriving an integral identity. It's noticed that

$$a(\omega) = \frac{2\theta(\omega)}{1 + \theta(\omega)} \text{Re}[1 - R(\omega)] = \eta(\omega)\text{Re}[1 - R(\omega)], \qquad (5)$$

The scaling factor $\eta(\omega) = 2\theta(\omega)/[1 + \theta(\omega)]$ is not merely a mathematical trick; rather, it captures the essence of energy flow in the waveguide system if we interpreted it by Thévenin's theorem as shown by Fig. 3.

First, we block the surface of the absorber to obtain equivalent source pressure, which is $2P_I$ obviously. Fahy terms it as block field pressure [30]. The source impedance viewing from the blocked loading is $Z_0 = 1$, namely, the impedance of waveguide. By reimplementing the loading, which is the sound absorber with impedance of $Z$, we get the following circuit representation for the 1D sound absorption problem.

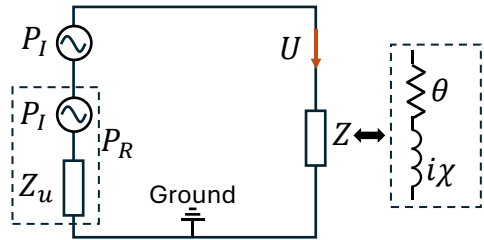

Figure 3: Equivalent circuit model for sound absorption problem in 1D waveguide.

Through such an electro-acoustic analogy, the governing equation reads

$$2P_I = (Z + Z_0)U = (\theta + 1 + i\chi)U \qquad (6)$$

This equation can also be derived by velocity continuity condition $P_I/Z_0 - P_R/Z_0 = U$ with the relationship between the pushing pressure and velocity response at surface of the absorber $P_I + P_R =$

$ZU$. Eq. (6) suggests that we can interpret the model in Fig. 1(a) as a system where a forcing term ($2P_I$) drives both the waveguide ($Z_u$) and the sound absorber ($Z$) in tandem.

The ratio of sound intensity dissipated by the sound absorber to the incident intensity consumed hence is

$$\eta_1 = \frac{|U|^2 \theta}{\text{Re}(P_I U)} = \frac{2\theta}{1+\theta}. \tag{7}$$

which is exactly the scaling factor in Eq. (5). The real part of $1-R$ also has its physical meaning by rearranging it as $\text{Re}(1-R) = \text{Re}\left(\frac{P_I - P_R}{P_I}\right) = \text{Re}\left(\frac{P_I U^* - P_R U^*}{P_I U^*}\right) = \frac{I_{in} - I_{out}}{I_{in}}$, where $I_{in}$ is the intensity supplied to the system consisting of fluid loading and the sound absorber, $I_{out}$ intensity escaped form the system due to reflection wave $P_R$. Taking the real part of $1-R$ represents the total power consumed by the circuit system shown in Fig. 3. Consequently, $\eta_1 \text{Re}(1-R)$ gives the ratio of sound intensity dissipated by the absorber itself, corresponding to the sound absorption coefficient, as shown by Eq. (5). Constructing an integral entity using $1-R$ provides clear physical meaning, leading to the Herglotz function as

$$H_1(\omega) = i(1-R). \tag{8}$$

Due to all inputs and outputs being real-valued functions in time domain, the symmetricity $H(\omega) = -H^*(-\omega)$ satisfied, where the subscript "$*$" means taking conjugate. It should be noted that there are no poles in upper half complex $\omega$ plane for $1-R$ of a passive system, hence there are no needs to introduce Blaschke product [1,19] or any other ancillary functions [2,11] to remove poles. The inequality presented in [1,2] arises from the use of ancillary functions. Ideally, our method has the potential to derive sum rules in the form of equalities, which is a key advantage of the current approach.

Combinations of Eq. (3), Eq. (5) and Eq. (8) yield

$$\frac{2}{\pi}\int_0^{+\infty} \frac{1+\theta(\omega)}{2\theta(\omega)} \frac{a(\omega)}{\omega^2} d\omega = \lim_{\varepsilon \to 0^+} \lim_{\delta \to 0^+} \frac{2}{\pi}\int_\varepsilon^{\varepsilon^{-1}} \frac{\text{Im}[H_1(\omega + i\delta)]}{\omega^2} d\omega, \tag{9a}$$

and

$$\frac{2}{\pi}\int_0^{+\infty} \frac{1+\theta(\omega)}{2\theta(\omega)} \frac{a(\omega)}{\omega^0} d\omega = \lim_{\varepsilon \to 0^+} \lim_{\delta \to 0^+} \frac{2}{\pi}\int_\varepsilon^{\varepsilon^{-1}} \frac{\text{Im}[H_1(\omega + i\delta)]}{\omega^0} d\omega. \tag{9b}$$

The asymptotic expansions of $H_1(\omega)$ at $\omega \to 0$ and $\omega \to \infty$ are

$$H_1(\omega)|_{\omega \to 0} = \frac{2}{K_e(0)}\omega^1 + O(\omega^2), \tag{10a}$$

and

$$H_1(\omega)|_{\omega \to \infty} = -\frac{2}{M_e(\infty)}\omega^{-1} + O(\omega^{-2}), \tag{10b}$$

respectively. It means

$$a_{-1} = 0, \quad a_1 = \frac{2}{K_e(0)} = \frac{L_e}{c_0}, \tag{11a}$$

and

$$b_{-1} = -\frac{2}{M_e(\infty)}, \quad b_1 = 0, \tag{11b}$$

compare to Eq. (3a) and Eq. (3b). $K_e(0) = c_0/L_e$ is effective stiffness of a composite structure, where $L_e$ is the effective depth of the absorber. We can use $K_e(0) = \lim_{\omega \to 0} i\omega \cot kL_e$ to convert $K_e(0)$ to $L_e$. $M_e(\infty)$ is the effective dynamic mass of a structure. Combinations of Eqs. (3), Eqs. (9) and Eqs. (11) we get the integral results.

$$\int_0^{+\infty} \frac{1+\theta(\omega)}{2\theta(\omega)} \frac{a(\omega)}{\omega^0} d\omega = \frac{\pi}{M_e(\infty)}, \quad (12a)$$

and

$$\int_0^{+\infty} \frac{1+\theta(\lambda)}{2\theta(\lambda)} a(\lambda) d\lambda = \frac{2\pi^2 c_0}{K_e(0)}. \quad (12b)$$

Before we reach final results, we should first address the scaling factor. In common situations, the damping $\theta(\omega)$ is frequency dependent, for example, couple resonators [31] and other composites [32]. In such consideration, we can use $\frac{1+\theta_{max}}{2\theta_{max}} \leq \frac{1+\theta(\omega)}{2\theta(\omega)} \leq \frac{1+\theta_{min}}{2\theta_{min}}$ to simplify the integrals to get the sum rules, which are

$$\frac{2\theta_{min}}{1+\theta_{min}} \frac{\pi}{M_e(\infty)} \leq \int_0^{+\infty} \alpha(\omega) d\omega \leq \frac{2\theta_{max}}{1+\theta_{max}} \frac{\pi}{M_e(\infty)}, \quad (13a)$$

and

$$\frac{2\theta_{min}}{1+\theta_{min}} \frac{2\pi^2 c_0}{K_e(0)} \leq \int_0^{+\infty} \alpha(\lambda) d\lambda \leq \frac{2\theta_{max}}{1+\theta_{max}} \frac{2\pi^2 c_0}{K_e(0)}. \quad (13b)$$

Note that $\int_0^{+\infty} \alpha(\lambda) d\lambda = 2\pi c_0 \int_0^{+\infty} \frac{\alpha(\omega)}{\omega^2} d\omega$ is used to obtain Eq. 13(b).

The first sum rule, Eq. (13a), is without the logarithmic operator and frequency weighting. This sum rule provides both lower and upper bounds, which are directly determined by the effective dynamic mass and the scaling factor. It suggests that a composite structure consisting of Lorentz resonators should minimize its dynamic mass to raise the upper bound; thus, a zero-mass system would be ideal. Interestingly, when examining impedance formulas for porous materials, mass terms are notably absent [33]. This indicates that porous materials, which absorb noises well across a broad frequency range, are among the best sound absorbers in theory. However, practical noise sources often concentrate their energy within a limited bandwidth, particularly at low frequencies, while porous materials are typically effective at higher frequencies. This mismatch means that porous materials may not be the best option for low-frequency noise control.

The second sum rule is with the integrant of $a(\lambda)$, or equivalently, $a(\omega)/\omega^2$. Unlike in [1], the logarithmic operator is removed here, allowing the sum rule to directly indicate the amount of energy that can be absorbed. An interesting observation is that the upper bound provided by Eq. (13b), when considering $\frac{\theta_{max}}{1+\theta_{max}} \to 1$, is essentially identical to that presented in Eq. (3.5) of [1]. This is not coincidental but is determined by $\lim_{\lambda \to \infty} -\ln[1-\alpha(\lambda)] \to \alpha(\lambda)$. The difference lies in the factor $\theta_{max}/(1+\theta_{max})$. This indicates that by knowing the maximum damping of a composite, a more compact bound can be found. For example, if we know $\theta_{max} = 1$, the upper bound given by Eq. (13b) is halved.

For both sum rules in Eq. (13a) and Eq. (13b), if we know the range of $\theta(\omega)$, we can fast conclude the upper bound and lower bound of the sound absorption performance of an absorber. This would be a very useful tool in practical design. To further illustrate the implications of Eqs. (13), we can apply mean value theorem to them to obtain sum rules for sound absorption in an arbitrary frequency band:

$$\bar{\alpha}_\omega \Delta\omega \leq \frac{2\theta_{max}}{1+\theta_{max}} \frac{\pi}{M_e(\infty)} \leq \frac{2\pi}{M_e(\infty)} \tag{14a}$$

$$\bar{\alpha}_\lambda \Delta\lambda \leq \int_0^{+\infty} \alpha(\lambda)d\lambda \leq \frac{2\theta_{max}}{1+\theta_{max}} \frac{2\pi^2 c_0}{K_e(0)} \leq \frac{4\pi^2 c_0}{K_e(0)}, \tag{14b}$$

where $\Delta\omega = \omega_2 - \omega_1$, $\Delta\lambda = \lambda_2 - \lambda_1$. $\bar{\alpha}_\omega = \frac{\int_{\omega_1}^{\omega_2}\alpha(\omega)\omega}{\omega_2-\omega_1}$ and $\bar{\alpha}_\lambda = \frac{\int_{\lambda_1}^{\lambda_2}\alpha(\lambda)d\lambda}{\lambda_2-\lambda_1}$ is the average sound absorption coefficient in the frequency interval of $[\omega_1\ \omega_2]$ and wavelength interval of $[\lambda_1\ \lambda_2]$, respectively.

An alternative form of Equation (14b) provides a direct relationship between the average sound absorption coefficient, the thickness-to-wavelength ratio, and the relative bandwidth of an absorber:

$$\bar{\alpha}_\lambda \cdot \bar{\omega} \leq \frac{\theta_{max}}{1+\theta_{max}} 4\pi^2 L_\lambda \leq 4\pi^2 L_\lambda, \tag{14c}$$

where $\bar{\omega} = \Delta\omega/\omega_c = (\omega_1 - \omega_2)/\sqrt{\omega_2\omega_1}$ is the relative bandwidth, $\omega_c = \sqrt{\omega_1\omega_2}$ is the center frequency of the frequency band of interest and $L_\lambda = L_e/\lambda_c$ is the thickness-to-wavelength ratio. Note that $L_e$ is the equivalent thickness by calculating $K(0) = \lim_{\omega\to 0} i\omega Z(\omega)$. Eq. (14c) can be used to predict the maximum average absorption coefficient once a targeted frequency band and thickness of sound absorber are fixed.

## 3.2 Comparison to existing sum rules

As discussed in the introduction, sum rules have been extensively studied since the research on circuit networks by Bode and Fano [3,7]. These works laid the foundation for subsequent developments, including sum rules for electromagnetic wave absorption [11] and sound absorption [2]. With the advancement of applying Herglotz functions to study various physical systems [19], more sum rules have been discovered, highlighting the fundamental constraints governing these systems.

A key question is whether different bounds are consistent and how they differ from one another. Comparing integrals over an infinite wavelength range is impractical, as few are concerned with absorption over such a range. Instead, we can compare the sum rule in Equation (14c) with the bounds derived by Meng *et. al* and Yang *et. al* [1,2] within a finite bandwidth. To investigate, Meng's and Yang's bound is rewritten in a limited band form, which is

$$\bar{\alpha} \leq 1 - e^{-\frac{4\pi^2 L_\lambda}{\bar{\omega}}} \tag{15}$$

Noted that $\gamma$ factor in [1] is excluded due to no porous materials are used in the comparisons. When $L_\lambda/\bar{\omega}$ is small enough, which means a broadband absorber working in deep-subwavelength scale, Eq. (14c) and Eq. (15) are identical once the Taylor expansion is applied to the right-hand-side term of Eq. (15) around $L_\lambda/\bar{\omega} \to 0$. For other situations, we show them in the following figure.

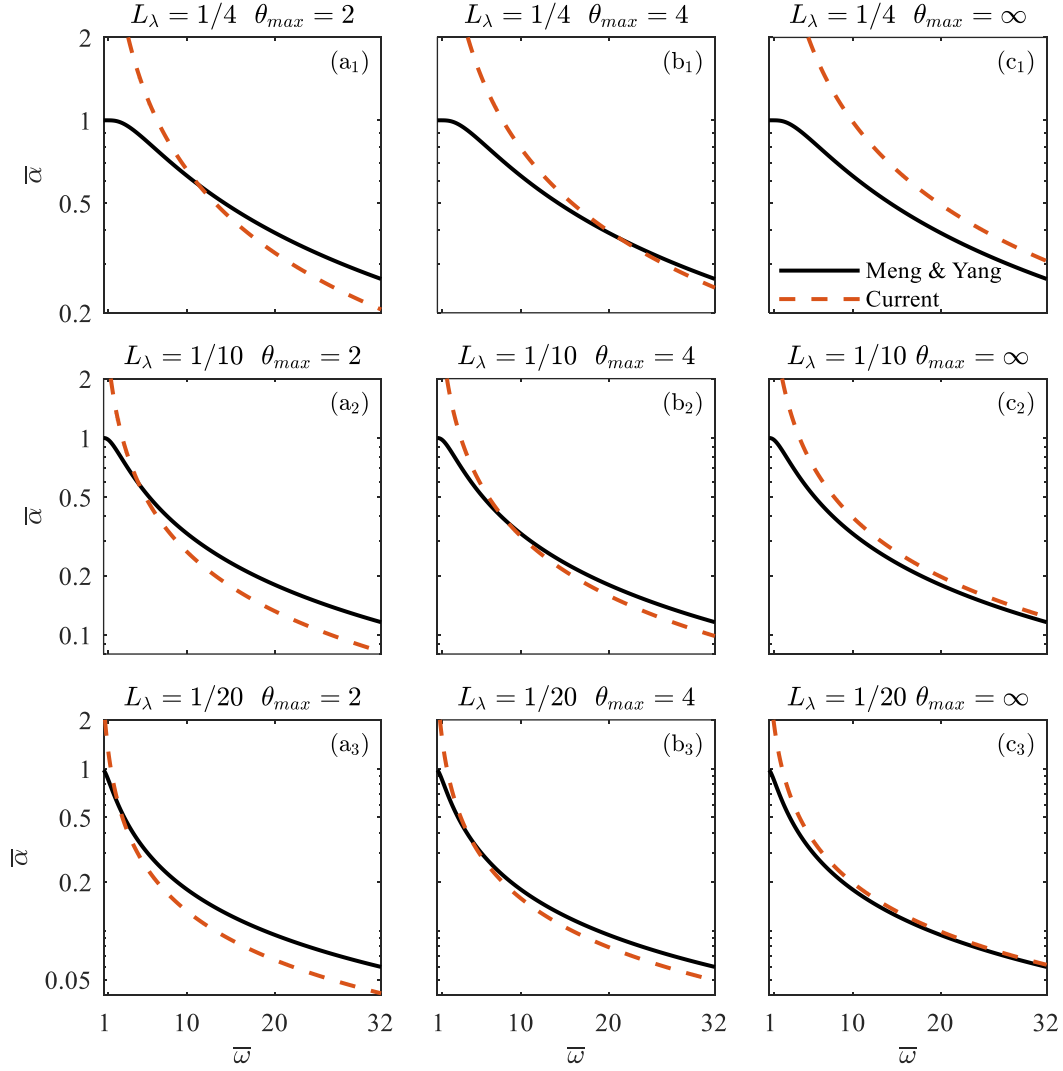

Figure 4: Compare predicted average sound absorption coefficient predicted by the sum rule of Eq. (11c) with different depth-to-wavelength ratio ($L_\lambda$) and maximum damping ($\theta_{max}$) to existing sum rules. (a$_1$-a$_3$) $\theta_{max} = 2$ and $L_\lambda = \frac{1}{4}, \frac{1}{10}, \frac{1}{20}$. (b$_1$-b$_3$) $\theta_{max} = 4$ and $L_\lambda = \frac{1}{4}, \frac{1}{10}, \frac{1}{20}$. (c$_1$-c$_3$) $\theta_{max} \to \infty$ and $L_\lambda = \frac{1}{4}, \frac{1}{10}, \frac{1}{20}$. All figure panels share the same legends while dash curves represent the current sum rule, and solid curves are for Meng et. al [1] and Yang et. at [2].

Fig. 4 shows that, for a compact sound absorber, Eq. (14c) generally provides a compacter bound for the average sound absorption coefficient if the maximum damping of the absorber is well estimated, as illustrated in Fig. 3(a3) and Fig. 3(b3). For a bulky absorber, the bound given by Eq. (14c) is too loose, and Eq. (15) is preferred. In practical designs, if the damping spectrum of a deep-subwavelength and broadband absorber is well understood, Eq.(14c) can quickly provide the upper bound for the average sound absorption coefficient. Additionally, with a targeted sound absorption coefficient and spectrum, the required depth can be rapidly predicted using Equation (14c). A win-win choice is to assemble them together in engineering practice.

## 3.3 The waterbed effect

Equation (14c) illustrates the "waterbed effect" in sound absorption, a concept first introduced by Bode in control theory [5]. This principle states that the capacity of a linear time-invariant (LTI) control system has an upper limit. When control performance is improved in certain frequency bands, the control performance at remaining frequency bands will deteriorate. This spectral balance resembles a waterbed with a fixed volume of water.

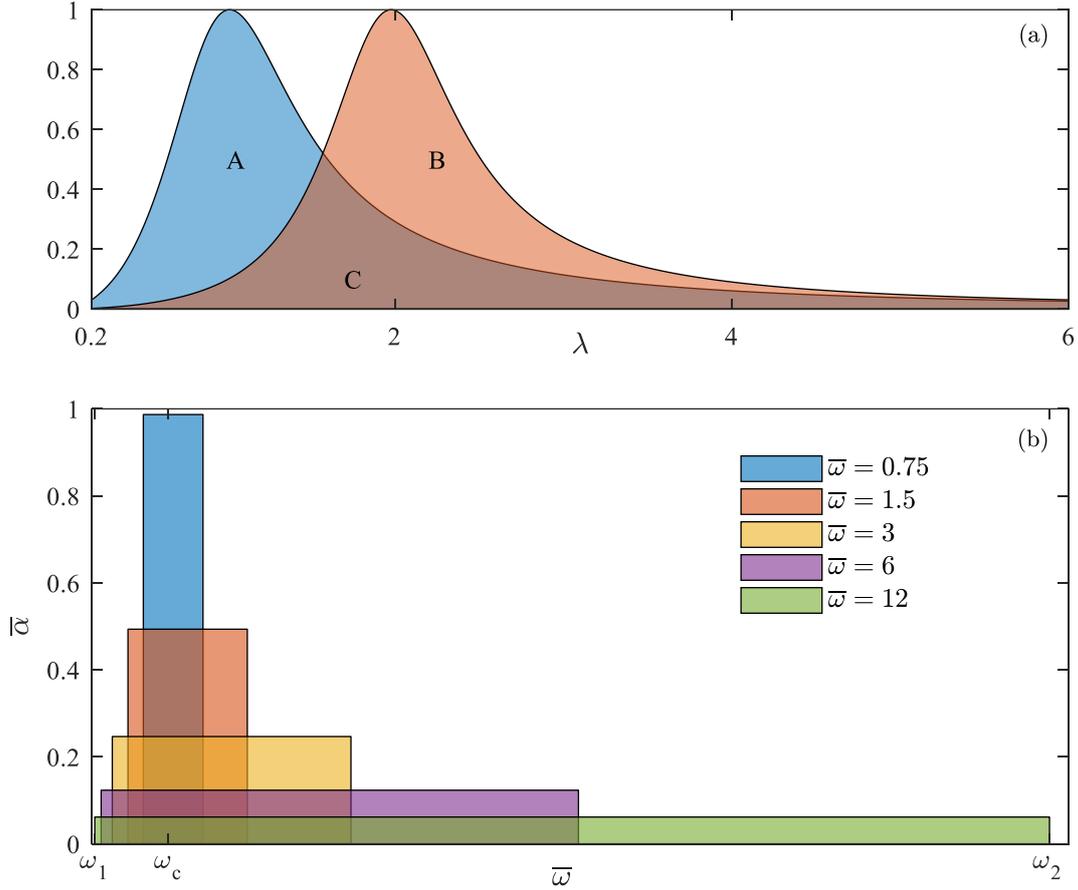

Figure 5: Demonstration of waterbed effect for linear time-invariant sound absorption structure. $L_\lambda = 1/20$ is used. (a) The waterbed effect of single resonator. (b) The compromise between the average sound absorption coefficient and the absorptive bandwidth.

Specifically, in the context of sound absorption, when the thickness of a sound absorber is fixed, improving sound absorption in some wavelength intervals will degrade it in others. Let us express inequality (11b) in a discrete form:

$$\Delta\lambda \sum_{1}^{N} \bar{\alpha}_n \leq \frac{4\pi^2 c_0}{K}, n = 1,2,3 \ldots \quad (16)$$

where $\Delta\lambda = (\lambda_2 - \lambda_1)/N$ is a small interval of wavelength. Obviously, $\bar{\alpha}_m > \bar{\alpha}$ leads $\sum_1^n \bar{\alpha}_{n,n\neq m} < \bar{\alpha}$. Another situation occurs when the average sound absorption coefficient is increased, causing the absorptive bandwidth (or wavelength range) to become narrower. The waterbed effect is illustrated in the following figure.

In Fig. 5(a), we design two resonators with identical stiffness and damping ($\theta =1$) but different dynamic masses. The results show that the shaded areas of the sound absorption spectra for both resonators are identical, indicating that adjusting the dynamic mass does not enhance the sound absorption coefficient in the wavelength domain. When shifting to the wavelength domain, it becomes evident that tuning the effective stiffness also does not improve the upper bound of the overall sound absorption coefficient. Fig. 5(b) demonstrates that as the targeted bandwidth ($\bar{\omega}$) broadens, the maximum average sound absorption coefficient decreases within this frequency interval.

**3.4 Typical cases**

Here, we use examples to illustrate the above analysis. In the field of metamaterials research, many artificial structures are designed by arranging multiple resonators in parallel, cascading, or hybrid configurations, with the aim of creating sound absorbers with broader effective bandwidths and higher sound absorption coefficients. We demonstrate that if these designs are passive and linear time-invariant system, they will be subject to sum rules and waterbed effects discussed in the previous sections. The following cases examine spatially parallel and cascaded resonator arrays.

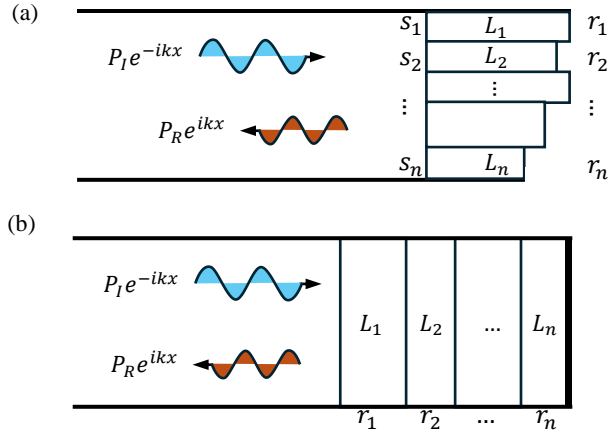

Figure 6: Typical resonator arrays for sound absorption purpose. (a) A sound absorber resembling resonators in a parallel manner. $n^{th}$ resonator is with an area ratio of $s_n$ and depth of $L_n$, and $\sum_{n=1}^{N} s_n = 1$. (b) A cascaded resonator array.

**(a) Spatial parallel absorber**

Without loss of generality, we use a spring-mass-damper system with a backing cavity to simulate a resonator. Such a system can represent most kinds of sound absorbents, such as MPPs, Helmholtz resonators, suspended diaphragms, membranes and so on. Noted the calculation cases

for a stretched membrane is shown in next section. The surface impedance for the parallelled array is

$$Z_{para} = \frac{1}{\sum_{n=1}^{N} \frac{1}{Z_{pn}}}. \tag{17a}$$

The surface impedance of $n^{th}$ resonator is the addition of the impedance of the portion ($Z_n$) and its backing cavity ($Z_{cn}$):

$$Z_{pn} = Z_n + Z_{cn} = iM_n\omega + \theta_n + \frac{K_n}{i\omega} - i\cot kL_n. \tag{17b}$$

where $M_n$, $\theta_n$ and $K_n$ are dynamic mass, damping and stiffness per area of the partition forming the surface, respectively. With $\omega \to 0$, it's not hard to find that the effective static stiffness is

$$K_e(0) = \frac{c_0}{\sum_{n=1}^{N} s_n L_{e_n}} = \frac{c_0}{L_e}, \tag{18a}$$

where $L_{e-para} = \sum_{n=1}^{N} s_n L_{e_n}$ and

$$L_{e_n} = \frac{c_0}{K_n + \frac{c_0}{L_n}}. \tag{18b}$$

$L_{e_n}$ is the equivalent depth of $n^{th}$ resonator when accounting for the stiffness of the partition forming the surface. If the partition is permeable such as an MPP, $K_n$ is normally zero, and the effective depth is the actual depth, namely, $L_{e_n} = L_n$. Substituting Eq. (18a) into Eq. (13b) we get the sum rules for the paralleled array.

When evaluating the paralleled resonators in frequency domain, we should get the effective dynamic mass $M_e(\infty)$ to obtain the sum rules. In Stolz domain, as $\omega \to 0$, we have $\lim_{\omega \to \infty} -i\cot kL_n \sim 1$. This implies that at sufficiently high frequencies, a cavity appears infinitely long compared to the wavelength of the propagating sound, effectively behaving like air. Alternatively, we can us $Z_c = \frac{1+R}{1-R}$ to reach $\lim_{\omega \to \infty} -i\cot kL_n \sim 1$. According to the Kirchhoff-Stokes attenuation law, expressed as $P(0) = P(x)e^{-\Lambda(\omega)x}$ where $x$ is the transmission distance and $\Lambda(\omega) \sim \omega^2$ for air, the reflected sound becomes minimal compared to the incident sound after a propagation length, namely, $R \to 0$. Consequently, the cavity's impedance approaches that of air at very high frequencies. For example, a 4 MHz ultrasound attenuates at a rate of 30 dB per centimeter in air [34], and a cavity with 1-cm-depth would leads the reflection wave neglectable. The damping term of a Lorentz resonator, $\theta_n$, is thus finite and becomes negligible relative to

$i\omega M_n$ in extremely high frequencies. Neglecting the damping term in high frequency limit, we can safely determine the effective dynamics mass of a parallel absorber as

$$M_{e-para}(\infty) = \frac{1}{\sum_{n=1}^{N} \frac{1}{M_n}}. \tag{19}$$

Combination of Eq. (13a) and Eq. (19) the sum rule with frequency as variable can be obtained. Following using the model shown in Fig. 7(a) to demonstrate the sum rules of Eq. (13b) and waterbed effects numerically.

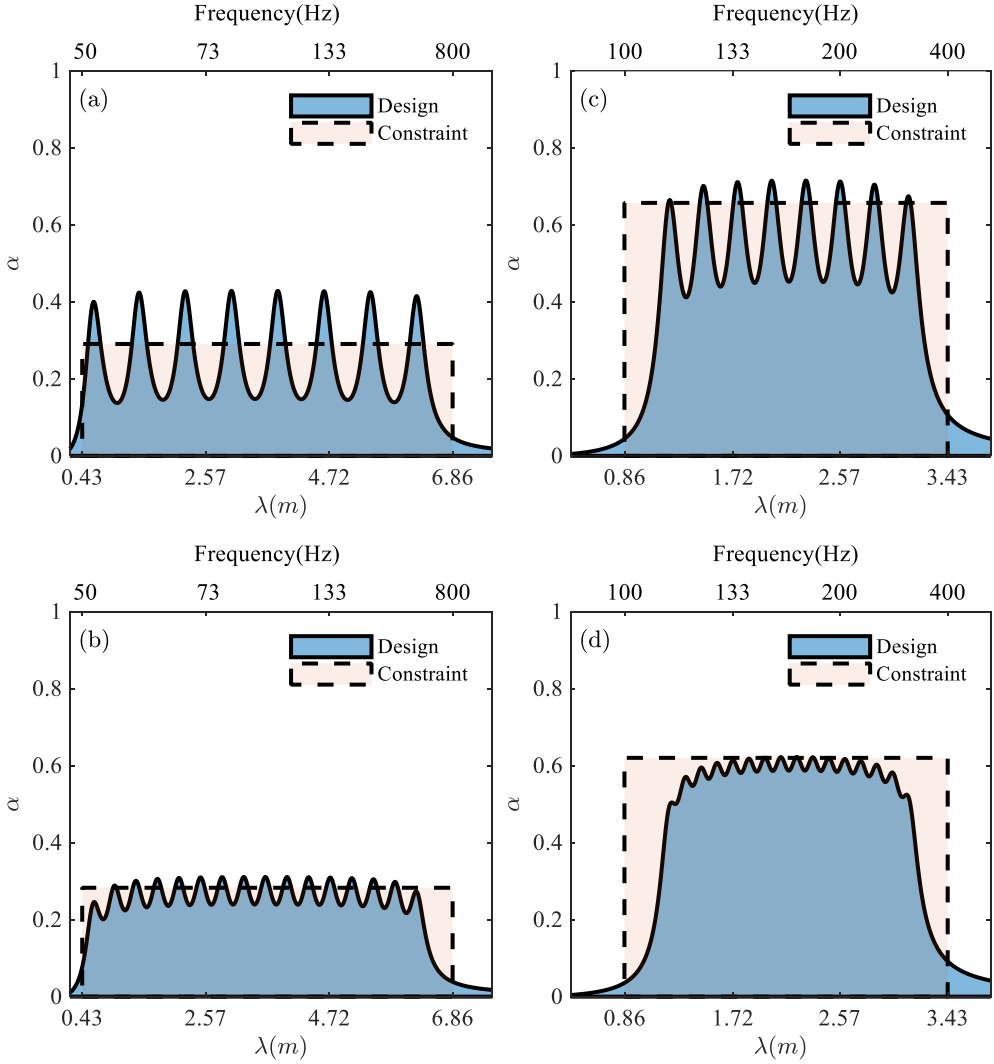

Fig. 7: Demonstration of the sum rule given by Eq. (13c) by sound absorbers consisting different numbers of resonators in paralleled and different targeted wavelength intervals. (a) Absorption spectrum of a sound absorber with 8 resonators and a targeted wavelength interval of $[\frac{c_0}{800\ Hz}\ \frac{c_0}{50\ Hz}]$. Intrinsic damping of each resonator is 1. The predicted bound by Eq. (13c) is shown by dash rectangular labeled by "constrain". (b) Case of 16 resonators and a targeted wavelength interval of $[\frac{c_0}{800\ Hz}\ \frac{c_0}{50\ Hz}]$. Intrinsic damping of each resonator is 1. (c) Case of 8 resonators and a targeted

wavelength interval of $[\frac{c_0}{400 \text{ Hz}} \frac{c_0}{100 \text{ Hz}}]$. Intrinsic damping of each resonator is 0.5. (d) Case for 16 resonators and a targeted wavelength interval of $[\frac{c_0}{400 \text{ Hz}} \frac{c_0}{100 \text{ Hz}}]$. Intrinsic damping of each resonator is 0.5. Depth of all resonators for all cases are 5 cm.

Fig. 7 illustrates cases for four deep-subwavelength absorbers, where the depth-to-wavelength ratio is $L/\lambda_c = 1/34.3$ for all cases. The relative bandwidth is $\overline{\omega} = 3.75$ for the cases shown in Fig. 7(a) and Fig. 7(b), and $\overline{\omega} = 1.5$ for the cases in Fig. 7(c) and Fig. 7(d). These cases clearly demonstrate that expanding the absorptive bandwidth results in a decrease in the average sound absorption coefficient, and vice versa. When sound absorption in certain wavelength ranges exceeds the average sound absorption, it results in reduced absorption at other wavelengths. Generally, the product of the sound absorption spectra and the targeted wavelength interval is constrained by the rectangular bounds given by the sum rule in Eq. (13c).

### (b) Cascading sound absorber

The impedance for a cascaded resonator array differs slightly. We can start with a two-resonator system, for which the impedance is

$$Z_{cas} = Z_1 + \frac{Z_{c1}(Z_2 + Z_{c2})}{Z_{c1} + (Z_2 + Z_{c2})}, \qquad (20a)$$

where $Z_{c1} = -i \cot k L_1$ and $Z_{c2} = -i \cot k L_2$ are the impedance of the first and the second cavities, respectively, and $Z_1 = M_1 i\omega + D_1 + K_1/i\omega$ and $Z_2 = M_2 i\omega + D_2 + K_2/i\omega$ are intrinsic specific acoustic impedance of first and second partitions. As $\omega \to 0$, the effective stiffness is

$$K_{e-cas}(0) = K_1 + \frac{K_{c1}(K_2 + K_{c2})}{K_{c1} + (K_2 + K_{c2})} \geq K_1 + \frac{K_{c1} K_{c2}}{K_{c1} + K_{c2}} = K_1 + K_{c1+2} \qquad (20b)$$

where $K_{c1+2} = c_0/(L_1 + L_2)$ is the static stiffness of the two cavities. Substitute Eq. (17b) into Eq. (10b) we can get the sum rule for two-layer system. It is here noted that a composite with multiple impermeable layers is not a good way to forming sound absorber, because the stiffness of the partition increases the static stiffness, which decreases the upper bound of the sound absorption given by the sum rule. With a number of layers, we can use iterative manner to get the static stiffness [35]. For *N*-layer the model shown in Fig. 6(b), we finally get

$$K_{e-cas}(0) \geq K_1 + K_{c1+2+..+N}. \qquad (21)$$

where $K_{c1+2+..+N} = \prod_1^N K_n / \sum_1^N K_n = c_0/\sum_1^N L_n$. Only all layers have zero stiffness holds the equality of Eq. (21).

Following the same processes, the effective dynamic mass

$$M_{e-cas}(\infty) \geq M_1, \tag{22}$$

which means the only the dynamics mass of the first layer is effectively. It's comprehensible, because at dynamic limit ($\omega \to \infty$), the transmitted sound through the first layer to the second layer is very small, and even will be drastically attenuated in the air. Thus, the following layers actually doesn't work in dynamic limits. In this stage, we can combine Eq. (22) with Eq. (13a) to obtain the sum rules with frequency as the variable.

## 4. Transmission problem

### 4.1 Theory

In the transmission model, we assume the thickness of the partition is much smaller compared to wavelengths of interests, as Fig. 1(b) shows, which is commonly true for most sound isolation structures. The downstream and upstream are viewed as fluid loading [30] with the normalized acoustics impedance of $Z_d$ and $Z_u$, respectively. The velocity response of a partition hence reads

$$U = \frac{2P_I}{Z} = \frac{2P_I}{Z_p + Z_u + Z_d} \tag{23}$$

where $Z_p$ is the normalized acoustics impedance of the partition itself. If the mediums in both sides of the waveguide are air, the upstream and downstream impedance are $Z_u = Z_d = 1$. Note that a partition can be with non-uniform velocity response across its surface, in this case, we can use the average impedance of the partition for $Z_p$ [26] in a 1D waveguide. The sound intensity transmission ratio hence is (chapter 5 in [30]):

$$\tau = \left|\frac{P_T U_T^*}{P_I U_I^*}\right| = \frac{4}{(\theta + 2)^2 + \chi^2}, \tag{24}$$

where $\theta = \text{Real}(Z_p)$ and $\chi = \text{Im}(Z_p)$. Again, we cannot use $1 - R$ to construct a Herglotz function. Noted that,

$$\text{Re}(1 - R) = \frac{2(\theta + 2)}{(\theta + 2)^2 + \chi^2} = \frac{\theta + 2}{2}\tau, \tag{25}$$

Therefore, the Horglotz function $H_1 = i(1 - R)$ can also be used to derive the sum rules for sound transmission. Once we replace the impedance of backing cavity in sound absorption analysis by the downstream fluid loading in this section, we get the sum rules for sound transmission in 1D waveguide directly. The sum rules are:

$$\int_0^\infty \frac{\theta + 2}{2} \tau(\omega) d\omega = \frac{\pi}{M_e(\infty)}, \tag{26a}$$

and

$$\int_0^\infty \frac{\theta + 2}{2} \tau(\lambda) d\lambda = \frac{2\pi^2 c_0}{K_e(0)}. \tag{26b}$$

For structures with frequency dependent damping, these two identities can be simplified as:

$$\frac{2}{\theta_{max}+2}\frac{\pi}{M_e(\infty)} \leq \int_0^\infty \tau(\omega)d\omega \leq \frac{2}{\theta_{min}+2}\frac{\pi}{M_e(\infty)} \tag{27a}$$

$$\frac{2}{(\theta_{max}+2)}\frac{2\pi^2 c_0}{K_e(0)} \leq \int_0^\infty \tau(\lambda)d\lambda \leq \frac{2}{\theta_{min}+2}\frac{2\pi^2 c_0}{K_e(0)} \tag{27b}$$

Again, we can write right-hand-side terms to limited band forms as

$$\bar{\tau}_\omega \Delta\omega \leq \frac{2\pi}{(\theta_{min}+2)M(\infty)} \leq \frac{2\pi}{M_e(\infty)}, \tag{28a}$$

$$\bar{\tau}_\lambda \bar{\omega} \leq \frac{4\pi^2 c_0}{(\theta_{min}+2)K(0)} \leq \frac{2\pi^2 c_0}{K_e(0)}, \tag{28b}$$

where $\bar{\tau}_\omega = \frac{\int_{\omega_1}^{\omega_2} \bar{\tau} d\omega}{\omega_2-\omega_1}$ and $\bar{\tau}_\lambda = \frac{\int_{\lambda_1}^{\lambda_2} \bar{\tau} d\lambda}{\lambda_2-\lambda_1}$. Therefore, the average sound transmission loss for arbitrary frequency and wavelength intervals, which are $\overline{TL} = 10\log_{10} 1/\bar{\tau}$, yield

$$\overline{TL}_\omega \geq \log_{10}\frac{(\theta_{min}+2)M_e(\infty)\Delta\omega}{4\pi} \geq \log_{10}\frac{M_e(\infty)\Delta\omega}{2\pi} \tag{29a}$$

$$\overline{TL}_\lambda \geq \frac{(\theta_{min}+2)K_e(0)}{4\pi^2 c_0} \geq \frac{\bar{\omega} K_e(0)}{2\pi^2 c_0} \tag{29b}$$

These results are different from [1], which gives upper bound of $TL$ (Eq. (3.25a) in ref. [1]), while current analysis gives lower bound of $TL$.

## 4.2 Cases study for sum rules for sound transmission loss

### (a) Tensioned Membrane

Here, we first show a case of a circular membrane as a partition for sound isolation. We can use Ingard's average impedance for a stretched membrane [26], which is

$$Z_M = -ikT_e \frac{J_0(k_M r)}{J_2(k_M r)} \tag{30}$$

Note that this impedance is slightly different from the formula in [26] due to the definition of sound propagation coordinates, and the fluid load $\rho_0 c_0$ in the original formula is normalized as 1 here. $T_e = \sigma/\rho_0$ is an equivalent thickness of a membrane, where $\sigma$ is areal density of the membrane. $k_M = \omega/c_M$ is the wave number of the membrane. $c_M = \sqrt{T/\sigma}$ is wave speed on the membrane, where $T$ is the tension force applied to the boundary of the circular membrane, and $r$ is the radius the membrane. $J_0$ and $J_2$ are zero- and second-order Bessel functions of the first kind, respectively.

The static and dynamic limits, or called the effective stiffness $K_e(0)$ and dynamic mass $M_e(\infty)$, of the membrane are obtained through deriving the asymptotic expansion of $Z_M$ as $\omega \to 0$ and $\omega \to \infty$, respectively.

$$K_e(0) = \lim_{\omega \to 0} i\omega \left[-ikL_M \frac{J_0(k_M r)}{J_2(k_M r)}\right] = \frac{8T}{\rho_0 c_0 r^2} \tag{31a}$$

For $\omega \to \infty$, we can first get the asymptotic expansions of $J_0(k_M r)$ and $J_2(k_M r)$ (page 118 in

[36] ), which is $J_n(ka) \sim \sqrt{\frac{2}{\pi x}} \cos\left(x - \frac{n\pi}{2} - \frac{\pi}{4}\right)$, and finally find

$$M_e(\infty) = \lim_{\omega \to \infty} \frac{1}{i\omega}\left[-ikL_M \frac{J_0(k_M r)}{J_2(k_M r)}\right] = \frac{\sigma}{\rho_0 c_0} \qquad (32b)$$

Eq. (32a) means that at extremely low frequencies, the membrane is moved like a piston and its effective stiffness is determined by the tension force and per area, which is reasonable in dynamics. At extremely high frequencies, the wavelength of the elastic wave of the membrane is very short. A very small structure damping would stop its travelling, considering which, segments of a membrane tend to move independently. Therefore, the dynamic mass term is proportional to the areal density. In this stage, we can substitute Eq. (32a) and Eq. (32b) into Eq. (27a) and Eq. (27b), respectively, to get the transmission sum rules for a membrane. We can also use Eq. (32a) and Eq. (32b) for sum rules of sound absorption problem to get the upper bound of membrane absorber.

**(b) Passive shunted loudspeaker**

Shunt techniques are recently often used to improve sound absorption in limited bandwidth such as series work from Lissek [37] and Zhang [38]. The impedance for a passive shunted loudspeaker is

$$Z_{ss} = \left[Mi\omega + \theta + \frac{K}{i\omega} + -i\cot kL_n\right] + \Delta Z(\omega), \qquad (33a)$$

where $\Delta Z$ is the electrically induced (EIA) acoustic impedance yielding

$$\Delta Z = \frac{(Bl)^2}{\rho_0 c_0 A(R_c + i\omega L_c + \Delta Z_e(\omega))} \qquad (33b)$$

$A$ is the effective area of loudspeaker diaphragm. $Bl$ is the force factor a moving-coil loudspeaker, which is the product of magnetic flux density (B) and effective length of a coil ($l$). $R_c$ and $L_c$ are the inherent resistance and inductance of a coil, respectively. $\Delta Z_e$ is the electrical impedance of the additional shunt circuit, which comprises of basic linear electrical elements, namely, resistors and capacitors and inductors. $\text{Re}(\Delta Z_e) + R_c > 0$ promises the passivity and stability.

No matter what the electrical elements are used and the network type for the additional circuit is, the circuit cannot alter the static stiffness $K_e(0)$ and dynamic mass $M_e(\infty)$ of the shunted loudspeaker. This point could be proof by contradiction. Supposing $\Delta Z = -\frac{(Bl)^2}{\rho_0 c_0 A \cdot i\omega\zeta}$ to induce a static negative stiffness of $\Delta K = -(Bl)^2/(\rho_0 c_0 A\zeta)$ to reduce the positive static stiffness of the acoustic part. Note that $\zeta$ is a real positive number. It will lead to $-1/\zeta i\omega = 1/(R_c + i\omega L_c + \Delta Z_e)$, which requires $R_c = 0$ and $i\omega L_c + \Delta Z_e = -i\omega\zeta$, resulting in $\Delta Z_e = -i\omega(\zeta + L_c)$. It means that we should collocate a circuit has zero resistance and negative inductance, which is unstable in fact. In the same way we can proof reduce the effective mass at high frequency limit is impossible for a stable system of a shunted loudspeaker. However, we can modify the damping of the shunted loudspeaker to improve the lower bound of sound transmission loss in arbitrary finite frequency band.

The static stiffness $K_e(0)$ and dynamic mass $M_e(\infty)$ of a shunted loudspeaker cannot be altered, regardless of the electrical elements used or the type of additional circuit network. This can be demonstrated by contradiction. Suppose $\Delta Z = -\frac{(Bl)^2}{\rho_0 c_0 A \cdot i\omega\zeta}$ is introduced to induce a static negative stiffness $\Delta K = -(Bl)^2/(\rho_0 c_0 A \zeta)$, thereby reducing the positive static stiffness of the acoustic part, where $\zeta$ is a real positive number. This would require $-1/\zeta i\omega = 1/(R_c + i\omega L_c + \Delta Z_e)$, which leads to $R_c = 0$ and $i\omega L_c + \Delta Z_e = -i\omega\zeta$. Consequently, $\Delta Z_e = -i\omega(\zeta + L_c)$, which implies the need for a circuit with zero resistance and negative inductance. Such a configuration is inherently unstable. Similarly, we can prove that reducing the effective mass at the high-frequency limit is impossible for a stable shunted loudspeaker system.

In another way, it is feasible to modify the damping of a shunted loudspeaker to enhance the lower bound of sound transmission loss within an arbitrary finite frequency band. When the net electrical reactance approaches zero and there are no resistive elements in the additional circuit, namely, $\text{Im}[i\omega L_c + \Delta Z_e(\omega)] \to 0$ and $\text{Re}[\Delta Z_e(\omega)] = 0$, the EIA acoustic damping can be expressed as:

$$\Delta\theta = \frac{(Bl)^2}{\rho_0 c_0 A R_c}. \tag{34a}$$

Therefore, the minimum of the total acoustic damping is

$$\theta_{min} = \theta + \frac{(Bl)^2}{\rho_0 c_0 A R_c}. \tag{34b}$$

Substituting Eq. (31b) into Eq. (27a) and Eq. (27b) we can get the lower bound of average sound transmission loss. By minimizing the coil resistance $R_c$, the lowest acoustic damping can be significantly increased, thereby improving lower bound of the average sound transmission loss.

## 5. Discussions and conclusions

### 5.1 Challenges for passive noise control

Since the sum rules for passive noise control were established [1,2,14,17,18,23], the design of passive noise control structures and composites appears to have reached a limit due to the fundamental constraints imposed by these rules.

When evaluating the logarithm reflection sensitivity from zero to infinite wavelength, the study in [1] highlights that porous materials are the best passive materials due to viscothermal losses, which introduce a factor of $\gamma$ that effectively enlarges the size of the inherent back cavity of a sound absorber. The analysis in this work (Eq. 11(a)) also demonstrates that when evaluating the sound absorption coefficient from zero to infinite frequency, porous materials remain the best option because they possess zero dynamic mass.

However, a porous material itself is not suitable for sound absorption in deep-subwavelength scale due to the cavity stiffness (or compressibility), which is inversely proportional to its depth, prevents low-frequency sound waves get into the porous materials in the absorber. For deep-subwavelength sound absorption with a limited volume, we normally need to induce dynamic masses term to improve the compressibility of the backing cavities in the vicinity of resonance frequency to resonating. For example, we often tune the aperture parameters of a MPP absorber to adjust the sound absorption spectrum in low frequencies, which is tuning the effective dynamic mass of MPP absorber actually. The dynamic mass, in turns impose sum rules to the absorber, as Eq. (11a) and Eq. (11b) depict.

However, porous materials themselves are not suitable for deep-subwavelength sound absorption because the cavity stiffness (or compressibility), which is inversely proportional to its depth, prevents low-frequency sound waves from penetrating into the porous material. For deep-subwavelength sound absorption with limited volume, it is typically necessary to introduce dynamic mass terms to enhance the compressibility of the backing cavities in the targeted frequency ranges. For instance, in a micro-perforated panel (MPP) absorber, we often adjust the aperture parameters to modify the low-frequency sound absorption spectrum, which essentially involves tuning the effective dynamic mass of the absorber. The introduction of dynamic mass, in turn, imposes sum rule constraints on the absorber, as depicted by Eq. (11a) and Eq. (11b).

In engineering practices, a high-frequency noise is relatively easy to be addressed, and the primary concern is usually noise control within a limited bandwidth at low frequencies. The long-standing research focus has been on achieving "broadband noise control by deep-subwavelength structures." Many efforts, both through conventional methods and modern metamaterial approaches, have been directed toward achieving broader absorption bandwidths, higher absorption coefficients, and smaller depth-to-wavelength ratios simultaneously. However, recent research and the analysis presented in this work show that these goals are inherently in conflict [1,2,17,18,23], and a trade-off that is encapsulated by the waterbed effect, as described in this study.

To overcome the constraints imposed by sum rules, a natural approach is to challenge the foundational assumptions on which these rules are built: causality, passivity, linearity, and time-invariance. Active systems and nonlinear structures are intuitive choices. The former can outperform passive systems by violating causality (through feedforward control) and passivity, but hardware costs, stability issues, and incompatibility with complex sound fields make it less favorable for many engineering applications. Nonlinear structures, on the other hand, are highly dependent on excitation amplitude, making consistent performance difficult to achieve. Recently, time-variant systems [39] employing temporal switching strategies have been reported to surpass the Bode-Fano-Rozanov bounds in electromagnetic wave absorption and transmission [40,41]. In acoustics, the potential comparable approach would be the switched shunted moving-coil [42,43], which has demonstrated the ability to convert sound frequencies in a linear manner.

However, linear time-invariant (LTI) systems are typically preferred in engineering practice due to their reliability and broad applicability. To mitigate the constraints imposed by sum rules by LTI systems, one potential strategy is the development of massless materials and structures, as well as reducing the static stiffness of a cavity with a fixed depth. As discussed earlier, porous materials serve as an example to achieve massless solution. To reduce the static stiffness is much difficult, which is determined by the depth of the cavity, namely, $K_e(0) = c_0/(K_0 + c_0/L)$. For deep-subwavelength structures, $c_0/L$ is the dominated term. There are many efforts try to create negative stiffness such as shunt techniques [38,44]. However, these works only demonstrate the reduce the effective stiffness at moderate frequencies instead of zero frequency. To reduce static stiffness to enlarge the effective depth of a cavity need further development.

## 5.2 Conclusions

In this work, we develop alternative sum rules for sound absorption and transmission problem of systems consisting of Lorentz resonators in a unidimensional waveguide, by removing the logarithmic operators and weighting functions. Based on these sum rules, the waterbed effect is demonstrated, revealing the inherent compromises among sound absorption performance, absorptive bandwidth and the thickness-to-wavelength ratio. Several cases of sound absorption are presented, demonstrating that the sum rules effectively predict the upper bound of the average sound absorption coefficient, particularly for deep-subwavelength structures designed to absorb sound in broadband at low-frequency regions.

**Acknowledgement:**

The authors wish to express their sincere gratitude to Dr. Yang Meng of CNRS, Le Mans University, for his insightful discussions and guidance throughout this work. Without his help we cannot finish this work. Special thanks to Prof. Lixi Huang of the University of Hong Kong for his insightful discussions and inspirational guidance. The work is supported by a National Natural Science Foundation of China (Grant No. 52105090), GuangDong Basic and Applied Basic Research Foundation (Grant No. 2023A1515110927), and National Natural Science Foundation of China (Grant No. 52405098).